\documentclass[aps,pra,twocolumn,showpacs,amsmath,floatfix]{revtex4-1}

\usepackage{graphicx}
\usepackage{color}
\usepackage{dcolumn}
\usepackage{bm}

\begin{document}

\title{Coupling approach for the realization of a $\mathcal{PT}$-symmetric
  potential for a Bose-Einstein condensate in a double well}

\author{Fabian Single}
\author{Holger Cartarius}
\email{Holger.Cartarius@itp1.uni-stuttgart.de}
\author{G\"unter Wunner}
\author{J\"org Main}

\affiliation{Institut f\"ur Theoretische Physik 1, Universit\"at Stuttgart,
  70550 Stuttgart, Germany}

\begin{abstract}
  We show how non-Hermitian potentials used to describe probability gain and
  loss in effective theories of open quantum systems can be achieved by a
  coupling of the system to an environment. We do this by coupling a
  Bose-Einstein condensate (BEC) trapped in an attractive double-$\delta$
  potential to a condensate fraction outside the double well. We investigate
  which requirements have to be imposed on possible environments with a linear
  coupling to the system. This information is used to determine an environment 
  required for stationary states of the BEC. To investigate the stability
  of the system we use fully numerical simulations of the dynamics. It turns
  out that the approach is viable and possible setups for the realization of
  a $\mathcal{PT}$-symmetric potential for a BEC are accessible.
  Vulnerabilities of the whole system to small perturbations can be attributed
  to the singular character of the simplified $\delta$-shaped potential used in
  our model.
\end{abstract}


\maketitle

\section{\label{sec:intro}Introduction}
In quantum mechanics the Hamiltonian is used to describe a system. It
determines the energy levels of all eigenstates and defines the time evolution
of a quantum state. It is common to work with Hermitian Hamiltonians to ensure
that the energies are real valued and the total probability is preserved in
time. These properties are no longer given for non-Hermitian Hamiltonians. They
appear in effective theories of open quantum systems \cite{Moiseyev2011a} and
may have imaginary potentials that represent probability sources or drains.

A way to obtain real eigenvalues in non-Hermitian systems is to balance the
total gain and loss. The simplest way to achieve this is by working with
$\mathcal{PT}$-symmetric systems, i.e.\ systems in which the
$\mathcal{PT}$-operator (consecutive execution of the time and parity
operators) commutes with the Hamiltonian $H$. For a standard Hamiltonian of the
form $H=p^2+V(x)$ this is equivalent to the condition
\begin{equation}
  \label{eq:pt}
  V(x) = V^*(-x) .
\end{equation}

Great interest in non-Hermitian $\mathcal{PT}$-symmetric systems was
generated by Bender and Boettcher in 1998 \cite{bub}. Since then,
$\mathcal{PT}$-symmetric systems have been the subject of theoretical studies
of quantum systems \cite{qm1,qm2,Mehri,Bender1999a}, quantum field theories
\cite{bub,Levai2002a,Bender2012a,Mannheim2013a,Bender2005a}, microwave cavities
\cite{Bittner2012a}, electronic devices \cite{Schindler2011a,Schindler2012a},
and in optical \cite{optic1,optic2} systems. The stationary Schr\"odinger
equation was solved for scattering solutions \cite{qm2} and bound states
\cite{Mehri}. Spectral singularities in $\mathcal{PT}$-symmetric potentials
\cite{Mostafazadeh2009a} turned out to be connected with the amplification of
waves \cite{Mostafazadeh2013a} and the lasing threshold
\cite{Mostafazadeh2013b}. Nonlinear $\mathcal{PT}$-symmetric quantum systems
have been discussed for BECs described in a two-mode approximation
\cite{Graefe08b,Graefe08a,Graefe10} and in model potentials
\cite{Musslimani08b}. The first experimental realization was achieved in
2009/2010, when two groups observed the expected properties in optical systems
\cite{opticPT1,opticPT2}. While these optical systems are analogues of quantum
mechanical systems, a real non-Hermitian \emph{quantum} system has yet to be
realized. A good candidate for such a realization is a BEC in a two-well
potential as suggested by Klaiman et al.\ \cite{Klaiman08a}.
This potential is extended with asymmetric imaginary parts, representing a
particle drain in one well and a particle source in the other.
Theoretical investigations revealed that this $\mathcal{PT}$-symmetric system
has observable eigenstates \cite{Graefe08b,Graefe08a,Graefe10,Car13a,%
  delta,ptsym,critpoint2,Haa14a,For14a}.
However, an experimental confirmation of these results cannot be made as long
as there is no way to realize a non-Hermitian potential.

We approach this issue with a simple model. To render it most instructive we
reduce the trap to two infinitesimally thin attractive $\delta$ potentials.
$\Delta$ functions have often been used to gain deeper insight with simple
solutions \cite{Mostafazadeh2006a,Jones2008a,Mostafazadeh2009a,qm2,Mehri,%
  Mayteevarunyoo2008a,Rapedius08a,Wit08a,Fassari2012a,Demiralp2005a,%
  Jones1999a,Ahmed2001a,Uncu2008a}.
By coupling the main system to a condensate outside the potential well we allow
the particle exchange caused by the non-Hermitian potentials. Both, the
coupling mechanism and the environment, will be constructed in a detailed
fashion.

We use the Gross-Pitaevskii equation (GPE) to describe a BEC of particles with
a short-ranged contact interaction. This equation describes a many-particle
system in a mean-field approximation \cite{gpe1,gpe2}. Due to the contact
interaction of the atoms in the dilute gas the GPE exhibits a nonlinearity
proportional to the modulus squared of the wave function. We note that the
presence of $\mathcal{PT}$-symmetry in nonlinear systems is not trivial since
the wave function appears in the Hamiltonian and thus also influences the
symmetries. The important properties of $\mathcal{PT}$-symmetric systems,
however, can still be observed when the eigenfunctions render the nonlinear
term in the GPE $\mathcal{PT}$-symmetric \cite{ptsym}. In our case this means
that the wave functions must have a symmetric modulus squared, which is
fulfilled for real chemical potentials \cite{Das13b}.

The basic outline of this article is as follows. We first summarize the
properties of the non-Hermitian double-$\delta$ potential and its eigenstates
in Sec.\ \ref{sec:ddp}. In Sec.\ \ref{sec:coupl} we introduce and discuss the
coupling approach mentioned above. The results are then applied in Sec.\
\ref{sec:3wave} and \ref{sec:2wave} to create different environments that can
sustain an eigenstate of the double-$\delta$ potential. Finally simulations of
the composite system of environment and stationary state are evaluated.
A discussion of the results and their relation to physical units is added
in Sec.\ \ref{sec:saa}.

\section{\label{sec:ddp}Non-Hermitian double-$\delta$
  potential}

To describe the condensate we use the Gross-Pitaevskii equation in the
dimensionless form introduced in \cite{delta}. Its time-dependent and 
time-independent versions read
\begin{subequations}
  \label{eq:gpe}
  \begin{align}
    i \frac{\partial}{\partial t} \psi &= \big[ -\Delta - g |\psi|^2 
    + V(x) \big]\psi, \\
    \label{eq:statgpe} \mu \psi &= \big[ -\Delta - g |\psi|^2 + V(x) \big]\psi,
  \end{align}
\end{subequations}
respectively, where $\mu$ is the chemical potential. An attractive interaction
is represented by a positive choice of the nonlinearity parameter $g$. When
the system is confined to one dimension, this parameter can be adjusted by
changing the trap frequencies confining the BEC \cite{transxy}. We use a
double-$\delta$ potential to represent the double well,
\begin{equation}
  V(x) = (V+i\Gamma)\delta(x-a) + (V-i\Gamma)\delta(x+a).
\end{equation}
This model has been used previously and proved to reproduce all important
features of double-well potentials with finite width \cite{delta}.
The system has three control parameters, i.e.\ the total distance between the
two wells $2a$, and the real and imaginary parts of the potential, $V$ and
$\Gamma$, respectively. In this paper the values $a=1.1$ and $V=-1.0$
are used, i.e.\ the unit of energy is chosen such that the real part of the
prefactor of the $\delta$ functions in the potential is normalized to unity.
Due to the non-Hermiticity parameter $\Gamma$ the potential models a particle
source at $x=a$ and a particle drain at $x=-a$. The nonlinearity $g$ in
the GPE \eqref{eq:gpe} can be seen as a fourth parameter.

\begin{figure}[tb]
  \includegraphics[width=\columnwidth]{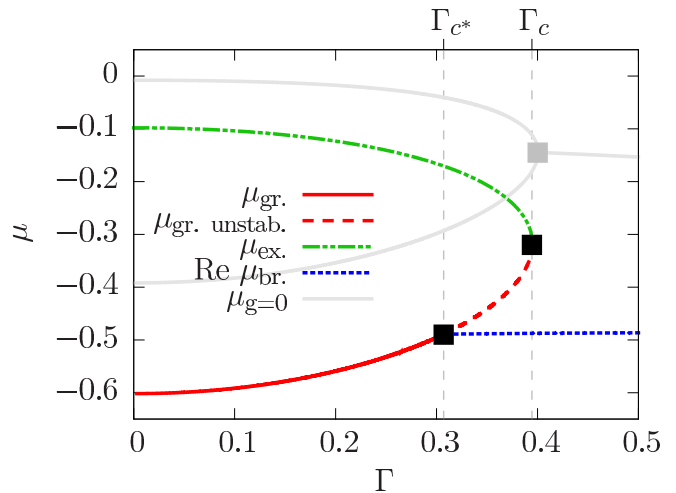}
  \caption{\label{fig:spectrum}(Color online) Eigenvalues $\mu$ of the
    double-$\delta$ potential in dependence of the non-Hermiticity parameter
    $\Gamma$. The spectrum is shown for $g=1.0$ (nonlinear case, strong and
    colored lines) and $g=0.0$ (linear case, light gray solid lines). Other
    parameters are fixed to $a=1.1$ and $V=-1.0$. The spectrum consists in both
    cases of a ground state $\mu_{\text{gr.}}$ and an excited state
    $\mu_{\text{ex.}}$ which merge in an exceptional point at $\Gamma_c$. Above
    a critical value $\Gamma_{c^*}$, where $\Gamma_{c^*} < \Gamma_c$ for
    $g\neq 0$, two $\mathcal{PT}$-broken solutions with complex and complex
    conjugate chemical potentials $\mu_{\text{br.}}$ appear, of which only the
    real part is shown in the figure. Ground state solutions between both
    critical points $\Gamma_c$ and $\Gamma_{c^*}$ are unstable 
    ($\mu_{\text{gr. unstab.}}$). All values are given in the dimensionless units
    introduced in Ref.\ \cite{delta}.}
\end{figure}

\begin{figure}[tb]
  \includegraphics[width=\columnwidth]{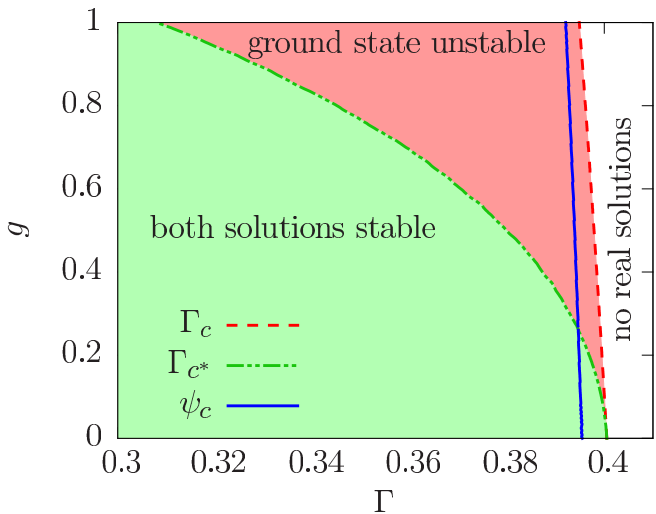}
  \caption{\label{fig:critpoint2}(Color online) Parameter space of the
    double-$\delta$ potential for fixed $a=1.1$ and $V=-1.0$. The dimensionless
    units introduced in Ref.\ \cite{delta} are used for all values. In the
    shaded areas stationary states with real eigenvalues exist. The lines
    separating the different areas show the location of the two critical points
    $\Gamma_c$ and $\Gamma_{c^*}$.
    Between these critical points the ground state solutions are unstable. The
    figure also shows where stationary states that satisfy relation
    \eqref{eq:phase3} in Sec. \ref{sec:one_bound} can be found ($\psi_c$).
    Many of these solutions are in the unstable regime (upper shaded area).}
\end{figure}

As found in \cite{Car13a} this system exhibits stationary states with real
eigenvalues for nonzero $g$ as long as the non-Hermiticity is not too strong.
Like in the linear regime ($g=0$), where analytic solutions are available, a
ground state and an excited state exist. Both are $\mathcal{PT}$-symmetric.
Figure \ref{fig:spectrum}
shows how the eigenvalues $\mu$ of these states depend on the parameter
$\Gamma$ when all other system parameters are fixed. With increasing
non-Hermiticity $\Gamma$ the energy gap between those states becomes
smaller. Eventually $\Gamma$ reaches a critical value $\Gamma_c$ at which the
eigenvalues as well as the wave functions of both states coalesce. Further
considerations show that this point fulfills all conditions of an exceptional
point \cite{spec}. Beyond this branch point the branches with real $\mu$
vanish. In the linear regime it is also the point where the
$\mathcal{PT}$-broken states appear. These states have eigenvalues
$\mu_{\text{br.}}$ with a nonvanishing imaginary part. For nonzero $g$ they branch
off at another critical point $\Gamma_{c^*}<\Gamma_c$. This critical point is
closely related to a stability change of the ground state, which can be shown
using a linear stability analysis \cite{Loe14a}. For values
$\Gamma<\Gamma_{c^*}$ the ground state turns out to be dynamically stable
with respect to small perturbations, above it is unstable. The stability of the
excited state does not depend on $\Gamma$. Figure \ref{fig:critpoint2}
shows the different stability regimes in parameter space together with the
location of the critical points.

\section{\label{sec:coupl}Coupling approach}

\begin{figure}[tb]
  \includegraphics[width=\columnwidth]{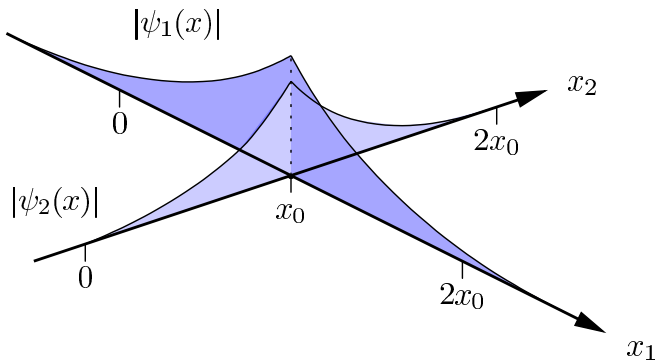}
  \caption{\label{fig:model}(Color online) Sketch of the geometric alignment of
    two one-dimensional stationary states $\psi_1(x)$ and $\psi_2(x)$. Both
    states intersect at $x_0$ in their coordinate system and interact at this
    point.}
\end{figure}

Non-Hermitian potentials represent particle sources or drains. Because
particles cannot just appear or vanish, such a potential can only be realized
by an environment that exchanges particles with the system. We model
such an environment using additional one-dimensional wave functions. In the
following discussion we will assume that an overlap of the stationary state
with its environment exists only in a $\delta$-shaped trap, cf.\ Fig.\
\ref{fig:model}.
Thus, the interaction between the wave functions is confined to a single
coupling or intersection point $x_0$. The sketch in Fig.\ \ref{fig:model}
indicates one possible geometry with which this can be obtained. Two
condensates with a confinement producing quasi one-dimensional waves are
arranged such that their extensions are orthogonal to each other and overlap in
one point. In the equations we reduce everything to one dimension and use the
same spatial coordinate for both wave functions to simplify the discussion.
Since the waves only interact at a single point $x_0$ (or two separated points
in the further discussion) the one-dimensional description can correctly
reproduce the physical situation shown in Fig. \ref{fig:model}.

In this section we discuss the form of the interaction and which requirements
an environment must satisfy to replace the non-Hermitian part of a single
$\delta$ potential. The results are then used to model environments for the
$\mathcal{PT}$-symmetric double-$\delta$ potential. Let $\psi_1(x)$ be a
stationary solution of the GPE \eqref{eq:statgpe} with the potential
$V_{\text{eff,1}}(x) = (V + i \Gamma)\delta(x-x_0)$ and the eigenvalue $\mu_1$.
We introduce another stationary state $\psi_2(x)$ acting as the environment.
This state is assumed to be exposed to the potential $V_{\text{eff,2}}(x) = 
(V+i\tilde\Gamma)\delta(x-x_0)$ and to have the chemical potential $\mu_2$.
Note that $\tilde\Gamma$ is not known yet, it will be derived later. The
parameters $\Gamma$ and $\tilde{\Gamma}$ are supposed to describe the flux
between both wave functions, i.e. an influx to one wave and an outflux from
the other. We introduce a linear coupling between both wave functions at the
intersection point. At the same time we remove all non-Hermitian contributions
from the potential, which leads to the following system of equations.
\begin{subequations}
  \label{eq:totsys1}
  \begin{align}
    \mu_1 \psi_1 &= \big[ -\Delta - g |\psi_1|^2 + V\delta(x-x_0) \big] \psi_1 
    \notag \\ & \quad + \gamma\psi_2\delta(x-x_0), \label{eq:totsysa}\\
    \mu_2 \psi_2 &= \big[ -\Delta - g |\psi_2|^2 + V\delta(x-x_0) \big] \psi_2 
    \notag \\ & \quad + \gamma\psi_1\delta(x-x_0). \label{eq:totsysb}
  \end{align}
\end{subequations}
The interaction strength is controlled by the coupling parameter $\gamma$.
One can easily see that the following condition must be satisfied to
reproduce $V_{\text{eff,1}}(x)$ for $\psi_1(x)$,
\begin{align}
  i\Gamma = \gamma \frac{\psi_2(x_0,t)}{\psi_1(x_0,t)}. \label{eq:condition}
\end{align}
From this we can derive several conditions that $\psi_2(x)$ has to fulfill.
By demanding that $\gamma$ is a real number we find a phase relation between
both wave functions at the interaction point,
\begin{equation}
  \arg \psi_2(x_0) - \arg \psi_1(x_0) = \pm \frac{\pi}{2} . \label{eq:phase}
\end{equation}
It is possible to choose $\gamma$ complex, but this would needlessly complicate
this relation and yield equivalent results. The real coupling parameter is used
to parametrize the strength of the mechanism. All phase information is
contained in the phase relations of the wave functions. Depending on the sign
chosen in Eq.\ \eqref{eq:phase} we obtain a different sign for $\gamma$,
\begin{equation}
  \label{eq:gamma}
  \gamma = \pm \Gamma \bigg| \frac{\psi_1(x_0)}{\psi_2(x_0)}\bigg| .
\end{equation}
From now on we choose the phase between both wave functions at the interaction
point such that $\gamma$ is positive. This simplifies the discussion below.
The implications of this choice will be discussed later when it actually yields
consequences. Finally Eqs.\ \eqref{eq:gamma} and \eqref{eq:totsysb} can be
used to calculate $\tilde{\Gamma}$ and the potential $V_{\text{eff,2}}(x)$ for
$\psi_2(x)$,
\begin{equation}
  \label{eq:veff}
  V_{\text{eff,2}}(x) = \Bigg(V - i \Gamma \bigg| \frac{\psi_1(x_0)}{\psi_2(x_0)} 
  \bigg|^2 \Bigg)\delta(x-x_0) .
\end{equation} 
The sign of the imaginary part originates from Eq.\ \eqref{eq:phase} and is
always negative independent of the sign chosen in that phase relation. It
should be emphasized that $V_{\text{eff,2}}(x)$ depends on $\psi_1(x_0)$. This
means that there is no universal environment because a change of $\psi_1(x)$
changes $V_{\text{eff,2}}$ which eventually changes $\psi_2(x)$. Since both states
are stationary, conditions \eqref{eq:condition} and \eqref{eq:phase} can only
be satisfied for all times if the states possess the same chemical potential, 
\begin{equation}
  \mu_2=\mu_1.
\end{equation}

Despite the asymmetry in $V_{\text{eff,1/2}}$ the total system \eqref{eq:totsys1}
is closed and Hermitian. One can easily check that the particle or probability
flux between both systems at the interaction point is opposite in sign and
equal in strength. A stationary state can be coupled to several others,
replacing an arbitrary number of non-Hermitian $\delta$ potentials. The wave
functions that are used to create this environment will hereafter be called
feeders. Using this method we can now construct an environment that can support
a stationary state of the non-Hermitian double-$\delta$ potential. This can be
done by either using two feeders, one for each well, or with only a single
feeder. Both methods will be investigated.

\section{\label{sec:3wave}Separate wave functions for
  each well}

\begin{figure}[tb]
  \includegraphics[width=\columnwidth]{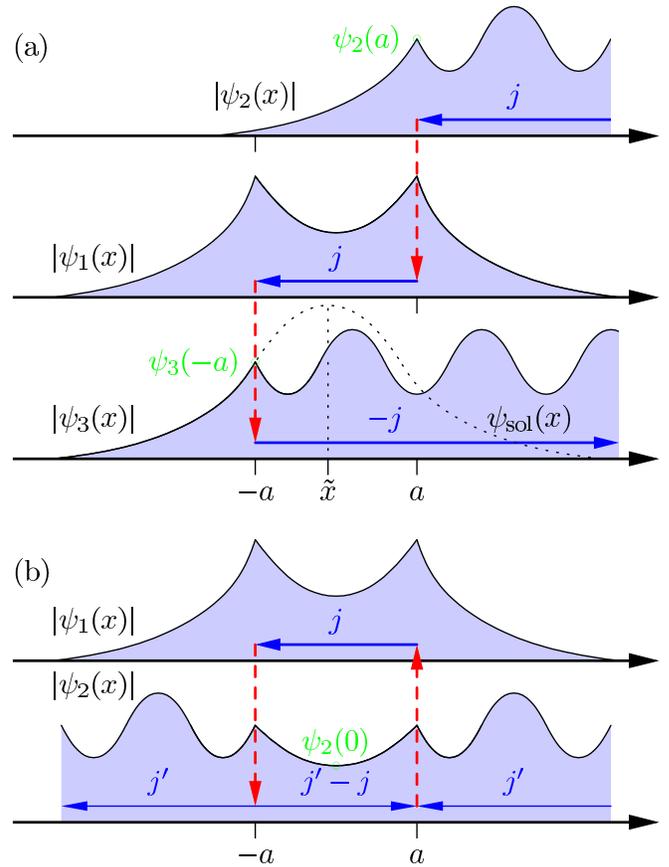}
  \caption{\label{fig:models}(Color online) Both figures show an environment
    supporting a stationary state $\psi_1$ of the double-$\delta$ potential.
    Particle currents $j$ are represented by solid blue arrows. Dashed red
    arrows represent a particle exchange between two wave functions. The free
    parameters for creating the environments are indicated as green dots. In
    (a) the environment consists of two feeders, particles are incoming via
    $\psi_2(x)$ and leaving via $\psi_3(x)$. In (b) the environment consists
    of only one wave function. The current it is carrying is partially
    redirected through $\psi_1(x)$. Figure (a) features a sketch how
    $\psi_3(x)$ is constructed with use of the soliton solution
    \eqref{eq:sol} (black dotted line). Its maximum $\tilde{x}$ is positioned
    such that the correct function value at the coupling point is achieved.}
\end{figure}

First we model an environment for a stationary solution of the double-$\delta$
potential $\psi_1(x)$ using one additional wave function for each $\delta$ peak.
The incoming beam $\psi_2(x)$ transports particles from a distant reservoir to
the well at $x=a$. Experimentally this could be realized with the use of
Bragg beams, outcoupled from a condensate fraction trapped in a third well not
part of the system investigated here \cite{bragg}. Another beam $\psi_3(x)$ is
used to remove particles at $x=-a$. The whole system is described by the
following set of equations.
\begin{subequations}
  \label{eq:3wave}
  \begin{align}
    \mu \psi_1 &= \big[ -\Delta - g|\psi_1|^2 + V\big(\delta(x-a) 
    + \delta(x+a)\big) \big] \psi_1 \notag \\
    &\hspace{.5cm}+ \gamma \psi_2 \delta(x-a) + \tilde{\gamma} \psi_3 
    \delta(x+a),\\
    \mu \psi_2 &= \big[ -\Delta - g|\psi_2|^2 + V \delta(x-a)\big] \psi_2 
    + \gamma \psi_1 \delta(x-a),\\
    \mu \psi_3 &= \big[ -\Delta - g|\psi_3|^2 + V \delta(x+a) \big] \psi_3 
    + \tilde{\gamma} \psi_1 \delta(x+a).
  \end{align}
\end{subequations}
Since the beam is directed it is reasonable to assume that $\psi_3(x)$ quickly
vanishes on one side of the interaction point, i.e.\ for $x\to-\infty$. We
shall make the same assumption for the incoming wave, which means that we expect
the incoupling to be perfect. To be precise, the whole flux of the incoming
beam is directed into $\psi_1(x)$. This will later simplify the stability
optimization and render the discussion more instructive. In principle any
non-perfect incoupling can be compensated by a larger amplitude of the incoming
wave. It also means that all three wave functions carry the same particle
current $j$. A sketch of the setup is depicted in Fig.\ \ref{fig:models}(a).

The only stationary solution of the potential-free GPE with $g > 0$ at hand
that satisfies $\lim_{x\to\pm\infty}\psi(x)=0$ is the bright soliton solution,
\begin{equation}
  \label{eq:sol}
  \psi_{\text{sol}}(x) = \sqrt{\frac{2 \kappa^2}{g}} \text{sech}
  \big(\kappa(x-\tilde{x})\big), \quad \kappa^2=-\mu .
\end{equation}
Only this function can describe the feeders $\psi_{2/3}(x)$ on the side of the 
interaction point on which they vanish. We construct $\psi_{2/3}(x)$ by choosing
the function value at the interaction points $\pm a$ ($+a$ for $\psi_2(x)$ and
$-a$ for $\psi_3(x)$). Consequently this value must be chosen from the interval
$(0,\sqrt{2 \kappa^2/g}]$. This is done by shifting the maximum $\tilde{x}$
of the soliton \eqref{eq:sol} appropriately. Recall that the stationary GPE
\eqref{eq:statgpe} is a second order differential equation. This means that a
function value $\psi_{2/3}(\pm a)$ and the value of the derivative
$\psi_{2/3}'(\pm a)$ are enough to determine $\psi_{2/3}(x)$ completely. Since we
require the form of the bright soliton solution even one of these values is
sufficient. The value of $\psi_{2/3}'(\pm a)$ can be derived from the value of
$\psi_{2/3}(\pm a)$ as is explained in appendix \ref{app:sol}. Thus, after
choosing the function value the feeders can be integrated to the left.
Integration to the right can be done after taking the potential
$V_{\text{eff,2/3}}(x)$ into account. This potential features a $\delta$ peak at
the interaction point, which causes a jump of the derivative. Figure
\ref{fig:models}(a) shows, how the soliton solution is used to partially
construct $\psi_3(x)$. The final construction step is to choose the global
phase of $\psi_{2/3}(x)$ such that the phase relation \eqref{eq:phase} is
fulfilled. In an experiment this phase could be changed by moving the double
well along the beams $\psi_{2/3}(x)$.

\begin{figure}[tb]
  \includegraphics[width=\columnwidth]{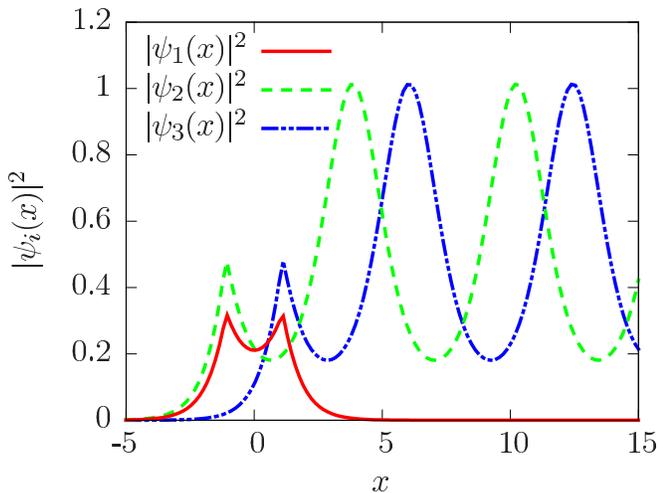}
  \caption{\label{fig:wave2}(Color online) Squared moduli of a stationary
    ground state of the double $\delta$ potential $\psi_1(x)$ and the two
    feeders $\psi_2(x)$ and $\psi_3(x)$. The numerical data was created with the
    parameters $a=1.1$, $g=1.0, \Gamma=0.1$ and $\psi_{2/3}(\pm a) =0.7$,
    which are given as all values in the figure in the dimensionless units
    introduced in Ref.\ \cite{delta}.}
\end{figure}

A system constructed in this way can be seen in Fig.\ \ref{fig:wave2},
where a stationary state of the double-$\delta$ potential (solid red line) and
both feeders (dashed and dashed-double-dotted lines) are plotted. Here both
feeders were created with the same function value at their interaction point.
Note that the squared moduli of both feeders are almost the same, they are only
translated in space. Apart from the translation the only difference between
both waves is the direction of the current they carry. The reason for this
symmetry originates from the construction of the feeders. On the right hand
side of both $\delta$ functions we require the soliton solution \eqref{eq:sol}.
The soliton wave $\psi_3$ has to possess the same function value at its
coupling point $x=a$ as the wave $\psi_2$ at $x=-a$ to ensure that the in- and
outfluxes are balanced. Due to the complex conjugate action of the $\delta$
functions the feeders evolve on the right hand side of their coupling points
with the same moduli but different phases.

Each environment of this kind has the free parameters $\psi_{2/3}(\pm a)$ which
can be picked such that certain properties of the feeders are optimized.
The most important property is their stability or their tolerance with respect
to perturbations. These feeders can be separated into two parts, one on each
side of the $\delta$ potential. On the left side the wave function is described
by the soliton solution while an unbound state of the potential-free GPE
describes it on the other side. The soliton solution itself is stable, as
observed in \cite{sol2,sol1}, however, the stability of the unbound solutions
can only be determined in a numerical time evolution, which we performed.

In the numerical simulations the waves possess a finite lifetime after which
they lose their stationary character due to numerical fluctuations in the
simulation. Stable solutions are characterized by damping these tiny
perturbations such that fluctuations are never observable. Unstable solutions
are destroyed since the perturbations increase with time. For details about
the simulation see appendix \ref{app:simulation}. We found that the environment
states are in many cases sufficiently stable to provide the in- and outflux of
the condensate fraction required for the implementation of imaginary
potentials. However, appropriate feeders have to be selected carefully. The
free parameters $\psi_{2/3}(\pm a)$ can be chosen in such a way that the highest
possible lifetime is achieved. This in particular requires $\psi_2(a) = 
\psi_3(-a)$ or $\tilde{\gamma} =\gamma$ because then both feeders have an
equivalent unbound part in terms of stability. With this optimization we found
environments that were capable of sustaining $\psi_1(x)$ for roughly $10$
reduced time units.

\section{\label{sec:2wave}Coupling to one wave function}
\subsection{Coupling to one unbound wave function}
It is also possible to create environments that only consist of one wave
function $\psi_2(x)$. This wave function then has two coupling points, one for
each non-Hermitian $\delta$ potential it has to replace. From geometrical
reasons alone an experimental realization of such a system is certainly more
challenging. We still perform this analysis because it is interesting from a
theoretical point of view. Is it possible to create an environment consisting
of only one wave function to replace a $\mathcal{PT}$-symmetric potential
which must allow for spatially separated source and drain effects?

The system is described by two coupled equations, viz.\
\begin{subequations}
  \label{eq:2wave}
  \begin{align}
    \mu \psi_1 &= \big[ -\Delta - g|\psi_1|^2 + V\big(\delta(x-a) 
    + \delta(x+a)\big) \big] \psi_1 \notag \\
    &\quad+ \gamma \big(\delta(x-a) + \delta(x+a)\big)\psi_2,\\
    \mu \psi_2 &= \big[ -\Delta - g|\psi_2|^2 + V\big(\delta(x-a) 
    + \delta(x+a)\big) \big] \psi_2 \notag \\
    &\quad+ \gamma \big(\delta(x-a) + \delta(x+a)\big)\psi_1.
\end{align}
\end{subequations}
As at the end of section \ref{sec:3wave} we shall use only one value of
$\gamma$ for both interaction points. Together with the symmetry of
$|\psi_1(x)|$ this imposes a condition on $\psi_2(x)$, viz.\
\begin{equation}
  \label{eq:abs2}
  |\psi_2(-a)|=|\psi_2(a)|,
\end{equation}
that in return renders $V_{\text{eff,2}}(x)$ $\mathcal{PT}$-symmetric, see
Eqs.\ \eqref{eq:gamma} and \eqref{eq:veff}. This means $\psi_2(x)$ has to be
a stationary state of the $\mathcal{PT}$-symmetric double-$\delta$ potential.
But these states are unbound unlike those discussed in section \ref{sec:ddp}.
Figure \ref{fig:models}(b) shows a sketch of an unbound feeder with two 
interaction points.

Before constructing the feeder we choose the state $\psi_1(x)$ for which we
create the environment. Now that $\psi_2(x)$ has two interaction points, the
phase relation \eqref{eq:phase} needs to be satisfied at both points, which is
only possible, if
\begin{align}
  \label{eq:phase2}
  \arg \psi_2(a) - \arg \psi_2(-a) = \arg\psi_1(a) -\arg\psi_1(-a) + \pi.
\end{align}
If two distinct values (or even one negative value) of $\gamma$ were allowed,
this relation would look different and lead to different environments.
However, a discussion of all possible environments would exceed the scope of
this work. Condition \eqref{eq:abs2} can easily be satisfied by demanding that
$|\psi_2(x)|$ has an extremal value at $x=0$ because solutions of the GPE
without potential are $\mathcal{PT}$-symmetric. We found that we can construct
such an extremal value by choosing $\psi_2(0)$ real and $\psi_2'(0)$ purely
imaginary. Both of these values were used as initial values for the numerical
integration of $\psi_2(x)$. Fulfilling relation \eqref{eq:phase2} is a more
difficult task which we tackled by a numerical analysis using both initial
values as variables. This analysis revealed that for any given $\psi_2(0)$
exactly one $\psi_2^{\prime}(0)$ exists such that $\psi_2(x)$ satisfies this
relation. Consequently, the choice of $\psi_2(0)$ already defines the whole
function. The final construction step is to set the global phase of
$\psi_2(x)$ in such a way that condition \eqref{eq:phase} is fulfilled.

\begin{figure}[tb]
  \includegraphics[width=\columnwidth]{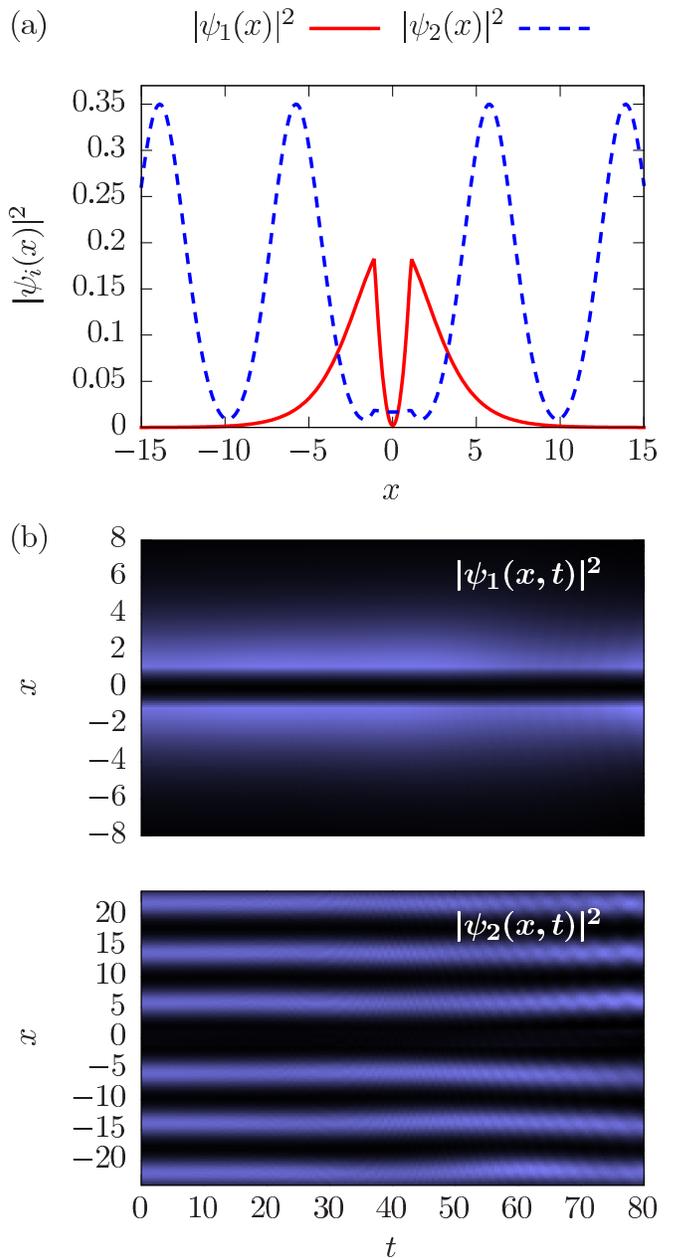}
  \caption{\label{fig:simu}(Color online) (a) Modulus squared of a stationary
    state of the double $\delta$ potential $\psi_1(x)$ and its feeder
    $\psi_2(x)$.
    (b) Simulated time evolution of both states. The system parameters in
    the dimensionless units from Ref.\ \cite{delta} used in the figure are
    $a=1.1$, $V=-1.0$, $\Gamma=0.1$ and $g=1.0$. Numerical perturbations
    cause both states to develop fluctuations, thus destroying the dynamics.
    They lose the stationary property at $t\approx45$. Bright colors represent
    high probability amplitudes, compare with figure (a) for absolute values.}
\end{figure}

Again there is one free parameter, $\psi_2(0)$, used in the creation of
$\psi_2(x)$. This parameter was used to optimize the stability of the system.
We were able to simulate systems with the excited $\mathcal{PT}$-symmetric
stationary state for roughly $45$ reduced time units. An example is shown in
Fig.\ \ref{fig:simu}.
The stationary $\mathcal{PT}$-symmetric state and its required feeder are
plotted in Fig.\ \ref{fig:simu}(a). The temporal evolution of both wave
functions can be found in Fig.\ \ref{fig:simu}(b), where density diagrams are
drawn. A long and stable numerical evolution was possible because the unbound
states that make up $\psi_2(x)$ could all be chosen to have a long lifetime,
i.e.\ they can be propagated numerically stable. Without the optimization the
system usually shows a numerical collapse as early as $t=5$. The environments
for the ground states are generally more unstable, resulting in lower numerical
lifetimes ($t \approx 10$) for these states until large fluctuations appear
and destroy the wave functions. To improve these lifetimes we suggest studying
such systems with a different choice of $a$ and $V$ or by the use of two
different signs of $\gamma$ for each interaction point.

\subsection{\label{sec:one_bound}Coupling to one bound wave
  function}
So far we assumed that all environments made from one wave function are
unbound. In this section we show that it is even possible to find
bound feeders. We found these solutions in an extensive numerical study,
but some properties can already be understood by a few simple considerations.
We know that both wave functions are stationary states of the double-$\delta$
potential. Also both $\psi_1(x)$ and $\psi_2(x)$ must have the same energy
eigenvalue. Since the spectrum of the double-$\delta$ potential is
non-degenerate (see Fig.\ \ref{fig:spectrum}) $\psi_2(x)$ can only be equal
to $\psi_1(x)$ or $\psi_1(-x)$. Only the latter approach is consistent with
relation \eqref{eq:phase2}. With this ansatz the phase relation turns into a
condition for $\psi_1(x)$, viz.\
\begin{equation}
  \label{eq:phase3}
  \arg \psi_1(a) - \arg \psi_1(-a) = \pi/2 .
\end{equation}
Only wave functions $\psi_1(x)$ that fulfill this condition can be coupled to
the bound environment. The position of these states in parameter space is
labeled $\psi_c$ in Fig.\ \ref{fig:critpoint2}. They are located
very close to the critical point $\Gamma_c$ and their non-Hermiticity parameter
is very large, with a value of $\Gamma \approx 0.39$. All of these states belong
to the ground state branch, which is why many of them are unstable (cf.\ Fig.\
\ref{fig:critpoint2}). This situation may change if the parameters $a$ and
$V$ are chosen differently. We note that the time evolution of these systems
could not be simulated correctly because of the large non-Hermiticity
parameter. Large values of $\Gamma$ caused errors in our simulations
originating from the singular character of the $\delta$ functions. This is
explained in more detail in appendix \ref{app:simulation}.

\section{\label{sec:saa}Discussion and outlook}
We successfully used a linear coupling scheme to replace non-Hermitian
$\delta$ potentials for BECs in a double-well setup. This coupling allowed a
particle exchange with an environment consisting of one or two additional
wave functions. Several methods to construct such wave functions for
stationary solutions of the double-$\delta$ potential were discussed. The
individual results were then tested in numerical simulations that maintained
the correct behavior for a certain time span. After this time numerical
perturbations caused unwanted dynamics, destroying the stationary character of
the system. The time this takes depends on the environment and the way it is
constructed.

For comparison with a realistic setup we present the time in SI-units for a
dilute BEC of $5000$ $^{\text{87}}$Rb atoms and a distance of $2a = 2.2\,\mu$m
between both potential wells. Then the longest stable simulation
was done with an unbound environment consisting of one wave function and
lasted for an equivalent of $123$\,ms. Other systems with two additional wave
functions could be simulated for up to $27$\,ms before the stationary
character of the system was lost. The simulations reveal that the mechanism
works well for a limited time.

We were able to simulate single unbound states for a longer time period than
the compound systems and found that the singular character of the $\delta$
potentials shortens this time considerably (cf.\ appendix
\ref{app:simulation}). A remedy could be a consideration of a system in which
potentials with finite width are used. This approach will be the next step in
extending the model. A more detailed stability analysis of the unbound
solutions of the potential-free GPE could also yield results that could be
used to improve this model further.

The results presented here are clearly only a first step towards the
possibility of realizing a $\mathcal{PT}$-symmetric potential with a coupling
of a BEC to condensate fractions outside the double well. The approach
provides an alternative to that suggested in Ref.\ \cite{4well1,4well2} where
the $\mathcal{PT}$-symmetric double-well was considered to be embedded in a
four-well structure. The results presented here are very encouraging since
they demonstrate the applicability of the approach. More realistic setups in
three dimensions with a specific description of the reservoir providing the
environment are necessary to gain more insight into possible experimental
realizations.

\appendix
\section{\label{app:simulation}Simulation}
The time evolution of all systems was simulated using the split operator method.
In this method the time evolution operator $e^{-i H \Delta t}$ is approximated
by the Baker-Campbell-Hausdorff formula,
\begin{align*}
  |\psi(t+\Delta t)\rangle &= e^{-i H \Delta t} |\psi(t)\rangle 
  = e^{-i\big[p^2 + V(x)\big]\Delta t}|\psi(t)\rangle \\
  &= e^{-ip^2\frac{\Delta t}{2}} e^{-iV(x)\Delta t} e^{-ip^2\frac{\Delta t}{2}} 
  e^{\mathcal{O}(\Delta t^3)}|\psi(t)\rangle .
\end{align*}
This approximation is suitable for a numerical evaluation because it is a
consecutive execution of operators that operate only in position \emph{or}
momentum space. A switch between both spaces via a fast Fourier transformation
reduces the application of the operators to simple multiplications.

In our simulations the wave function was discretized into $16384$ or $32786$
bins. The time step $\Delta t$ was chosen between $10^{-6}$ and $10^{-4}$
depending on the problem. We used a large number of bins to obtain a
high resolution in position space in the simulations containing $\delta$
functions. A $\delta$ potential $a\delta(x-x_0)$ was implemented by setting the
value of the potential at the bin closest to $x_0$ to $a/\Delta x$, where
$\Delta x$ is the resolution in position space. The nonlinear interaction term
of the GPE \eqref{eq:gpe} was implemented as a time-dependent potential.

We tested this simulation by evolving stationary states of the double-$\delta$
potential. These tests revealed that the method works well if the
non-Hermiticity parameter $\Gamma$ is small. Large values quickly caused small
numerical perturbations to become too large yielding wrong simulation results.
After extensive numerical tests we can relate these effects completely to the
singular character of the $\delta$ potentials. Combined with non-Hermitian terms
small errors then lead to an irreversible change of the norm instead of a
reversible redistribution of the energy. While $\delta$ potentials are easy to
handle in an analytical approach they are hard to implement in a discrete
numerical computation. Because of these purely numerical errors we restrict
our simulation to potentials with small $\Gamma$, usually $\Gamma=0.1$.

We also used the simulation to investigate the stability behavior of unbound
solutions of the potential-free GPE \eqref{eq:statgpe}. In these simulations
the wave functions eventually lose their stationary character indicating that
they are unstable. The time it took for this to happen depended on the
individual state and could be as small as $5$ reduced time units. For some
states it was independent of the simulation accuracy used. We classified
these states as unstable. For other states this time increased when $\Delta t$
was decreased or the number of bins was increased. These states were classified
as potentially stable and could be simulated for up to $70$ reduced time units.

\section{\label{app:sol}Soliton solution}
In this section we derive an expression of the derivative of the soliton
solution \eqref{eq:sol}.
\begin{align*}
  \frac{d}{dx} \psi_{\text{sol}}(x) &= - \sqrt{\frac{2 \kappa^2}{g}}
  \text{sech}\big(\kappa(x-x_0)\big) \text{tanh}\big(\kappa(x-x_0)\big)\\
  &= -\kappa \psi_{\text{sol}}(x) \text{tanh}\bigg[\text{arcsech}
  \bigg(\sqrt{\frac{g}{2\kappa^2}}\psi_{\text{sol}}(x)\bigg)\bigg]
\end{align*}
With the relation
\begin{equation}
  \tanh \text{arcsech}\hspace{.5mm} x = \pm \sqrt{1-x^2}
\end{equation}
we gain an expression for $\psi_{\text{sol}}'$ depending only on the function
value and $\kappa$,
\begin{equation}
  \label{eq:solder}
  \frac{d}{dx} \psi_{\text{sol}} = \kappa \psi_{\text{sol}} \sqrt{1-
    \frac{g \psi_{\text{sol}}^2}{2 \kappa^2}}.
\end{equation}
This formula is used in the construction of the feeders in Sec.\
\ref{sec:3wave}. It enables us to calculate the derivative of the wave
function with the function value which acts as a free parameter in the
construction of the feeders. Both values can then be used for the numerical
integration of the wave function. Note that there are two solutions because
of the square root. We only used the positive solution resulting in feeders
as depicted in Fig.\ \ref{fig:models}(a).


%

\end{document}